\documentclass[aps,prb,reprint,groupedaddress,amsfonts,showpacs,floatfix]{revtex4-1}

\usepackage{graphicx}
\usepackage{epstopdf}

\newcommand{\mrm}[1]{\mathrm{#1}}

\newcommand{\mbb}[1]{\mathbb{#1}}

\newcommand{\mc}[1]{\mathcal{#1}}

\newcommand{\fref}[1]{Fig.~\ref{#1}}
\newcommand{\sref}[1]{Sec.~\ref{#1}}
\newcommand{\tref}[1]{Table~\ref{#1}}

\newcommand{\rcite}[1]{Ref.~\onlinecite{#1}}

\newcommand{\yy}[2]{$y_1$=$#1$, $y_2$=$#2$}
\newcommand{\xn}{f_1}
\newcommand{\xnn}{f_2}

\begin{document}

\title{Classification of topological symmetry sectors on anyon rings}

\author{Robert N. C. Pfeifer}
\email[]{rpfeifer@perimeterinstitute.ca}
\affiliation{Department of Physics, The University of Queensland, Brisbane, QLD 4072, Australia}
\altaffiliation{Correspondence address: Perimeter Institute for Theoretical Physics, 31 Caroline St. N., Waterloo ON~~N2T 1G4, Canada}

\date{May 18, 2012}

\begin{abstract}
The golden chain with antiferromagnetic interaction is an anyonic system of particular interest as when all anyons are confined to the chain, it is 
readily stabilised against fluctuations away from criticality. 
However, additional local scaling operators have recently been identified on the disc which may give rise to relevant fluctuations in the presence of free charges. Motivated by these results for Fibonacci anyons, this paper presents a systematic method of identifying all topological sectors of local scaling operators for critical anyon rings of arbitrary winding number on surfaces of arbitrary genus, extending the original classification scheme proposed in Feiguin et al.~(2007). 
Using the new scheme 
it is then shown that for the golden chain, additional relevant scaling operators exist on the torus 
which are
equivalent to those detected on the disc, and which may disrupt the stability of the critical system.
Protection of criticality against 
perturbations generated by these additional scaling operators
can be
achieved by suppressing the exchange of charge between the anyon ring and 
the rest of the manifold.
\end{abstract}


\maketitle

\section{Introduction}

Fibonacci anyons are presently the subject of intense theoretical and experimental interest.\cite{feiguin2007,trebst2008,pfeifer2010,pfeifer2010a,nayak2008,xia2004,pan2008,kumar2010} 
They are the simplest non-Abelian anyons capable of implementing universal quantum computation through entirely topological means,\cite{freedman2002,kitaev2003,nayak2008} and it has previously been proposed that they may exist as quasiparticles in the $\nu=12/5$ %
fractional quantum Hall effect, as the non-Abelian component of the $k$=3 $Z_k$-parafermion Read-Reyazi state.\cite{read1999,xia2004,pan2008} With recent results \cite{kumar2010} %
adding further support to this hypothesis,
an understanding of the collective behaviour of these anyons is %
increasingly desireable.

The study of interacting Fibonacci anyons by numerical means was initiated in \citeyear{feiguin2007} by \citeauthor{feiguin2007},\cite{feiguin2007} who examined the behaviour of a ring of Fibonacci anyons on the torus with trivial winding, under the action of ferromagnetic (FM) and antiferromagnetic (AFM) nearest neighbour Hamiltonians. Both systems proved to be critical, and the authors introduced a system for classifying the resulting local scaling operators according to a topological symmetry of the system. In \rcite{pfeifer2010a} 
the same systems were re-examined on the disc, on which
it was shown %
that %
only one of these topological sectors survives. However, study of the same system on the disc using an anyonic version of the Multi-scale Entanglement Renormalisation Ansatz (MERA)\cite{konig2010,pfeifer2010,vidal2007,vidal2008a}  %
revealed the presence of additional local scaling operators which did not admit classification under the existing scheme.\cite{pfeifer2010}

In this paper I show that corresponding local scaling operators also exist on the torus, and introduce an extended classification system, valid on either surface, which %
expands the original classification of \rcite{feiguin2007} %
to incorporate the additional local scaling operators identified in \rcite{pfeifer2010}. This scheme is important as it provides, for the first time, a systematic method of identifying all topological sectors of a non-self-intersecting anyon ring with arbitrary winding number, on a surface of arbitrary genus.
In addition, for rings of Fibonacci anyons, it is shown that the topological protection of criticality described in \rcite{feiguin2007} only extends to the new sectors detected in \rcite{pfeifer2010} (and classified in this paper) provided
charges are prevented from entering or leaving the anyon ring.

\section{Construction of local scaling operators}

\subsection{On the disc}

\begin{figure}
\includegraphics[width=246.0pt]{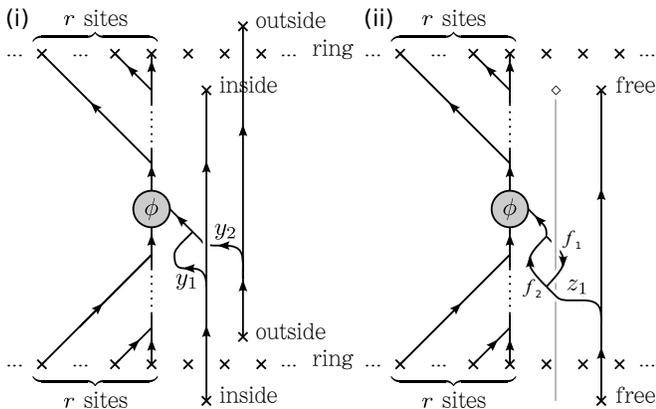}
\caption{Graphical representations of %
$r$-site local operators in the notation of \protect{\rcite{pfeifer2010}}, 
(i)~on a ring of anyons on the disc, interacting with regions of free charge on either side of the ring, and (ii)~on a ring of anyons on the torus, interacting with a free charge elsewhere on the surface of the torus. The symbol $\times$ denotes a site on the anyon ring or a free charge, and the symbol $\diamond$ that the fusion tree in~(ii) encircles a non-trivial cycle of the torus perpendicular to the anyon ring. Consequently the loop in~(ii) cannot be eliminated from the description of the operator (see also \protect{\rcite{pfeifer2010a}}).
\label{fig:locscalop}}
\end{figure}

In \rcite{pfeifer2010}, the authors directly studied the local scaling operator spectrum of an infinite chain of Fibonacci anyons on the disc under the action of ferromagnetic (FM) and antiferromagentic (AFM) Hamiltonians. The operators were constructed, and their scaling dimensions computed, using an anyonic Scale-Invariant MERA.\cite{pfeifer2009} 
In this work the authors allowed for the possibility of free charges in the regions to either side of the chain, %
and considered scaling operators on the ring which could interact with these charges.
Written in a basis in the form of \fref{fig:locscalop}(i), these operators admit an interpretation in which they transfer charges of $y_1$ and $y_2$ between these external regions and the anyon ring.
As the Fibonacci anyon model has only one non-trivial charge, this approach yields four sectors of local scaling operators, characterised by $y_1$ and $y_2$ taking on the different possible combinations of %
$\mbb{I}$ and $\tau$.

The scaling dimensions in the \yy{\mbb{I}}{\tau} and \yy{\tau}{\mbb{I}} sectors are seen to be identical, and as the regions connecting to $y_1$ and $y_2$ in \fref{fig:locscalop}(i) may be interchanged by means of an orientation-reversing map, it follows that the associated local scaling operators are equivalent up to a sign on their conformal spins.

\begin{figure}
\includegraphics[width=246.0pt]{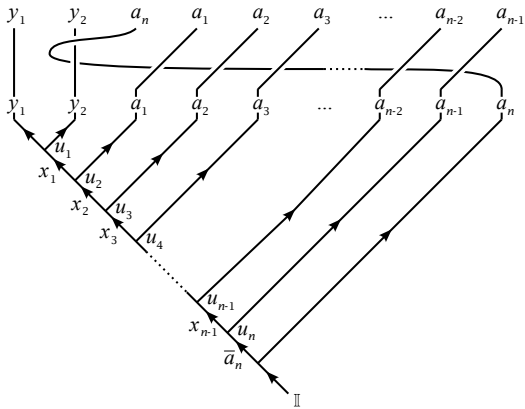}
\caption{Translation operator acting on a ring of anyons on the disc in the presence of two free charges $y_1$ and $y_2$.\label{fig:disctranslop}}
\end{figure}

It is well-known that the scaling dimensions of a critical system may also be obtained by exactly diagonalising the Hamiltonian, with the energy spectrum (appropriately shifted and rescaled) yielding the scaling dimensions in the thermodynamic limit.\cite{cardy1996,difrancesco1997} However, previous study of Fibonacci anyons on the disc using this technique has been restricted to the \yy{\mbb{I}}{\mbb{I}} sector.\cite{pfeifer2010a} To obtain scaling dimensions for non-trivial $y_1$ and $y_2$, it is necessary to exactly diagonalise the Hamiltonian in the presence of non-trivial free charges outside and inside the ring, taking on values $y_1$ and $y_2$ respectively. The translation operator then takes the form of \fref{fig:disctranslop}, and results for %
\yy{\tau}{\mbb{I}} are given in \tref{tab:chiralsectors}. 
\begin{table}[bp]%
\caption{Energy spectra, associated scaling dimensions, and conformal spins, %
for a ring of (i)~26 and (ii)~27 Fibonacci anyons on the disc with AFM interactions in the presence of a free $\tau$ charge. Energies have been shifted and rescaled to match their conformal field theory %
assignments, and conformal spins are determined from the momentum.
Note that Table~\protect{\ref{tab:chiralsectors}}(i) reproduces the scaling dimensions computed using the Scale-Invariant MERA in \protect{\rcite{pfeifer2010}}.\label{tab:chiralsectors}}
~\\
\begin{tabular}{|c|cc|c|}
\hline
\phantom{i}(i)~Energy~\phantom{(ii)} & \multicolumn{2}{c|}{Scaling} & Conformal spin \\
$n$=26 & \multicolumn{2}{c|}{dimensions} & for $y_1$=$\tau$, $y_2$=$\mbb{I}$\\
\hline\hline
0.4750 & ~~~~0.4750~~~~ & $(\frac{7}{16},\frac{3}{80})$ & $\frac{2}{5}$ \\
0.6000 & 0.6000 & $(0,\frac{3}{5})$ & $-\frac{3}{5}$ \\
1.4318 & 1.4750 & $(\frac{7}{16}$,$\frac{3}{80}+1)$ & $\frac{2}{5}-1$ \\
1.4354 & 1.4750 & $(\frac{7}{16}+1$,$\frac{3}{80})$ & $\frac{2}{5}+1$ \\
1.5546 & 1.6000 & $(0,\frac{3}{5}+1)$ & $-\frac{3}{5}-1$ \\
1.5748 & 1.6000 & $(0+1,\frac{3}{5})$ & $-\frac{3}{5}+1$ \\
\hline
\end{tabular}
\\~\\~\\
\begin{tabular}{|c|c c|c|}
\hline
(ii)~Energy~\phantom{(ii)} & \multicolumn{2}{c|}{Scaling} & Conformal spin \\
$n$=27 & \multicolumn{2}{c|}{dimensions} & for $y_1$=$\tau$, $y_2$=$\mbb{I}$\\
\hline\hline
0.1000 & ~~~~0.1000~~~~ & ($0,\frac{1}{10}$) & $-\frac{1}{10}$ \\
0.4750 & 0.4750 & ($\frac{7}{16},\frac{3}{80}$) & $\frac{2}{5}$ \\
1.0862 & 1.1000 & ($0,\frac{1}{10}+1$) & $-\frac{1}{10}-1$ \\
1.4474 & 1.4750 & ($\frac{7}{16}+1,\frac{3}{80}$) & $\frac{2}{5}+1$ \\
1.4819 & 1.4750 & ($\frac{7}{16},\frac{3}{80}+1$) & $\frac{2}{5}-1$ \\
2.0162 & 2.1000 & ($0,\frac{1}{10}+2$) & $-\frac{1}{10}-2$ \\
\hline
\end{tabular}
\end{table}%

\subsection{On the torus}

\begin{figure}
\includegraphics[width=246.0pt]{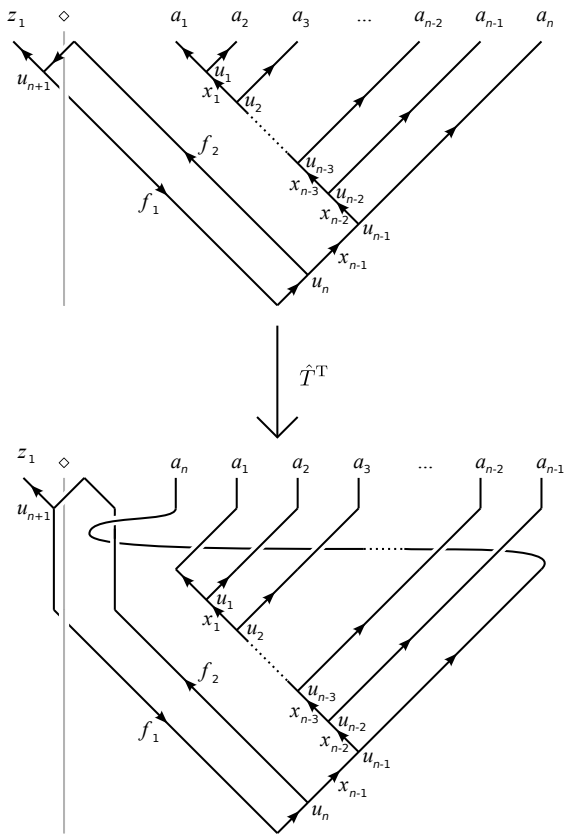}
\caption{Fusion tree and translation operator for a ring of anyons on the torus with trivial winding, in the presence of a free charge $z_1$. If the ring of anyons is assumed to lie on the larger circumference of the torus (assuming a circular cross-section), then the above fusion tree is constructed in the ``outside'' region of space. %
For $z_1\not=\mbb{I}$ there may no longer be a well-defined ``flux through the torus'', which is superceded instead by the three measurements $z_1$, $\xn$, and $\xnn$. The symbol $\diamond$ %
is an obstruction to contraction, and indicates that the labelled loop is topologically non-trivial, and cannot be eliminated from the description of the state.\protect{\cite{pfeifer2010a}}\label{fig:torustreewithfreecharge}
}
\end{figure}

In \rcite{pfeifer2010a} it was shown that a ring of anyons on the torus with a ``flux through the torus''\cite{feiguin2007} of $\xn$ and trivial winding number could be mapped to a ring of anyons on the equator of a sphere, with free charges at the north and south poles of $a_\mrm{N}=\overline{\xn}$ 
and $a_\mrm{S}=\xn$ 
respectively. It therefore should come as no surprise that the local scaling operators observed on the disc in the \yy{\mbb{I}}{\mbb{I}} and \yy{\tau}{\tau} sectors correspond to the operators observed on the torus having fluxes $\mbb{I}$ and $\tau$ respectively. However, it is also possible to obtain counterparts on the torus to the local scaling operators which inhabit the \yy{\mbb{I}}{\tau} and \yy{\tau}{\mbb{I}} sectors on the disc. This is achieved by introducing a free charge on the torus, and permitting local scaling operators to couple to this charge. Only one free charge need be considered as the ring of anyons does not bisect the torus, %
and so there is only one topologically distinct region capable of participating in charge transfers onto or off the anyon ring.
As with the disc, we can obtain the scaling dimensions and conformal spins of these scaling operators by computing the energy spectra and momenta of the system using exact diagonalisation, %
for an appropriate choice of translation operator and fusion tree (\fref{fig:torustreewithfreecharge}). 
When the free charge is non-trivial ($z_1$=$\tau$), the two sectors $\xn$=$\tau$, $\xnn$=$\mbb{I}$ and $\xn$=$\mbb{I}$, $\xnn$=$\tau$ yield scaling dimensions equivalent to the \yy{\tau}{\mbb{I}} and \yy{\mbb{I}}{\tau} 
sectors on the disc. However, we also obtain an additional set of scaling dimensions labelled by $z_1$=$\xn$=$\xnn$=$\tau$. The scaling dimensions in this sector are identical to those in the $z_1$=$\mbb{I}$, $\xn$=$\xnn$=$\tau$ sector, but because they have $z_1$=$\tau$, 
we may %
interpret them 
as operators 
mediating
a flow of charge onto or off the anyon ring. Thus we see that on the torus with trivial winding, there exist two local scaling operators for every one in the $y_1$=$y_2$=$\tau$ sector on the disc, both having $\xn$=$\xnn$=$\tau$, but distinguished by the value of $z_1$. %

\section{Classification of local scaling operators}

\subsection{Definition of classification system\label{sec:defclass}}

We now generalise the above discussion for a non-self-intersecting ring of an arbitrary species of anyons, with an arbitrary winding number, on a surface $S$
of arbitrary genus, at criticality. For the purposes of this classification, we shall allow that surface $S$ may have boundaries, but will distinguish between anyonic quasiparticles, found on the anyon ring or as free charges in the bulk, and punctures in surface $S$ which may also carry anyonic charge. In practice this distinction is synthetic as any anyon may be considered equivalent to a charged microscopic puncture in a quantum spin liquid on surface $S$.

To classify the local scaling operators %
on the ring, we begin by identifying each operator with a corresponding energy eigenstate having precisely one free charge, which may be trivial, in each topologically distinct region of the manifold (e.g. on the disc, the regions are inside and outside the anyon ring). In the event that there exist multiple punctures and/or quasiparticles within a given region of $S$, the relevant charge is recognised to be the total charge of all punctures and quasiparticles within that region. For purposes of this classification scheme, in each such region we replace
all punctures/quasiparticles with a single puncture whose boundary carries the total charge present within that region. Thus, for example, a ring of anyons on either the sphere or the disc is represented by a ring of anyons on the doubly-punctured sphere, with the ring dividing the sphere into two hemispheres, and one puncture in each hemisphere. The boundaries of these punctures then carry charges $y_1$ and $y_2$ respectively (which may be trivial).

For the eigenstate under consideration, we then identify %
an open disc $R\subset S$ %
covering all anyons on the ring but none of the free charges, and define a submanifold $P=S-R$. The classification of the operator is then given by specifying the value of a complete set of independent non-trivial charge measurements which may be performed on $P$ but which do not measure the total charge of all punctures on $P$ (or the dual of the total charge on $P$). We will denote such a set of measurements by $\mc{M}_P$. %
For rings of anyons on the sphere or disc, %
the natural choice is to 
measure
the two free charges $y_1$ and $y_2$. 
On a torus where the ring has trivial winding we 
may
measure %
the free charge $z_1$ and the flux through the two non-trivial cycles 
$\xn$ and $\xnn$ as %
labelled in \fref{fig:torustreewithfreecharge}. 
Similar sets of measurements may be constructed for surfaces of higher genus, and for anyon rings with non-trivial winding numbers.

Due to the relationship between the torus and the disc described in \rcite{pfeifer2010a}, when the winding number of the ring on the torus is zero we observe the following corollaries:
\begin{enumerate}
\item For any sector on the torus having specified flux values $\xn$ and $\xnn$, the sector on the disc having charges $y_1=\overline{\xn}$, $y_2=\xnn$ will yield the same set of scaling dimensions. %
\item If we denote by $N^{c}_{ab}$ the multiplicity tensor for the fusion rules of the anyon model, then for any sector on the disc having free charges $y_1$ and $y_2$, there will be as many equivalent sectors on the torus as there are values of $z_1$ for which $N^{z_1}_{\overline{y_1}y_2}$ is non-zero. 
These sectors %
correspond to the different admissible values of charge $z_1$ on the torus for fluxes $f_1$=$\overline{y_1}$, $f_2$=$y_2$.
\item As there always exists some charge $a$ such that $a\in \overline{y_1}\times y_2$, there is always at least one sector on the torus corresponding to a given sector on the disc.
\item Consequently, when all sectors are taken into account, the same values of scaling dimensions will always be obtained on both the torus and the disc. %
\end{enumerate}

\subsection{Completeness of classification system}

It is readily seen that the above classification system incorporates all possible sectors: If one denotes by $\mathcal{M}_R$ a complete set of independent charge measurements on the disc $R$, then no matter what the topology of the manifold $S$ or the arrangement of the anyon ring, by definition the set of measurements $\mc{M}_P\cup \mc{M}_R$ provides complete information about the anyonic system (subject to the condition, imposed in \sref{sec:defclass}, that for an anyonic system where multiple free charges or punctures exist within a single topologically distinct area of $S$, these are replaced by a single puncture carrying the equivalent total charge; this replacement is admissible as local scaling operators on the ring do not interact with any structure within these regions, but only with the total charge). Likewise, by definition the set of measurements $\mc{M}_R$ contains only information which may be obtained from local measurements performed on the collection of anyons on the ring, and indeed contains all such information. As $\mc{M}_P\cap\mc{M}_R$ is empty, it follows that $\mc{M}_P$ contains only information about the system which is not locally encoded on the anyons of the ring (and hence which may be considered topological, as it relates only to the genus of manifold $S$, and to the total free charge on the boundaries of $S$), and further, that it contains \emph{all} such topological information about the system. Thus a classification system based on a set of measurements $\mc{M}_P$ is necessarily complete. Furthermore, by construction there necessarily always exists at least one valid choice of $\mc{M}_P$ and $\mc{M}_R$ (though $\mc{M}_R$ will be trivial if the ``ring'' is defined to contain no anyons).

Finally, by the linear independence of the measurements in $\mc{M}_P$ and by the completeness of the information which they provide about the topological degrees of freedom of the system, it follows that regardless of the actual choice of measurements $\mc{M}_P$ (e.g. one may exchange a measurement for its dual), the resulting classification of local scaling operators into topological sectors is unique.
\\

\section{Implications for the stability of Fibonacci anyon rings}

An important result of \rcite{feiguin2007} was that, provided translation invariance is enforced, the AFM golden chain on the torus without free charges is protected against perturbations which are ``relevant'' in the renormalisation group sense. This result was obtained by observing that the operator $\hat Y$ (described in \rcite{feiguin2007}) classifies states according to the flux through the torus, which is a topological symmetry of the system. The only relevant local scaling operator in the same sector as the ground state is $\sigma$, a local scaling operator with scaling dimension $\frac{7}{8}$, which is suppressed by maintaining translation invariance. This result carries forward to the present classification system, where once again $\sigma$ is the only relevant local scaling operator in the same sector as the ground state.

Furthermore, recall that %
local scaling operators having non-trivial $y_1$, $y_2$, or $z_1$
may be interpreted as mediating a transfer of charge onto or off the anyon ring (Fig.~\ref{fig:locscalop}).
For any local scaling operator whose scaling dimension is found exclusively in a sector with at least one non-trivial free charge, fluctuations in the Hamiltonian %
which are generated by this operator will be suppressed if we 
prohibit this transfer of charges onto or off the ring.
For the AFM golden chain, all relevant scaling dimensions in %
sectors with non-trivial free charges %
are indeed found only in those sectors, and hence
if (and only if) the transfer of charge onto and off the ring is %
suppressed
will 
the topological protection of criticality %
described in \rcite{feiguin2007} %
be 
preserved.

\begin{acknowledgments}
The author acknowledges the support of the Australian Research Council (%
APA). This research was supported in part by the Perimeter Institute for Theoretical Physics. The author also thanks the Ontario Ministry of Research and Innovation ERA for financial support.
\end{acknowledgments}

\end{document}